\documentclass[doublecol]{epl2} 
\usepackage{amssymb}
\usepackage{graphicx}
\usepackage{dcolumn}
\usepackage{amsmath}
\usepackage{bm}
\usepackage{epsfig}

\def\defi{{\buildrel \;def\; \over =}}
\newcommand{\be}{\begin{equation}}
\newcommand{\ee}{\end{equation}}

\newcommand{\mediaT}[1]{\left\langle #1 \right\rangle}

\newcommand{\media}[1]{\langle #1 \rangle}

\def\mod{ \mathop{\rm mod} }

\title{Critical phenomena on heterogeneous small-world networks}
\shorttitle{Critical phenomena on heterogeneous small-world networks}
\author{Massimo Ostilli \inst{1,2} and Jos\'e F. F. Mendes \inst{1}} 
\institute{                    
  \inst{1} Departamento de F{\'\i}sica and I3N, Universidade de Aveiro, 3810-193 Aveiro, Portugal \\
  \inst{2} Statistical Mechanics and Complexity Center (SMC), INFM-CNR SMC, Rome, Italy.
}

\abstract{
We consider critical phenomena on heterogeneous small-world networks
having a scale-free character but also arbitrary short-loops.
After deriving the self-consistent equation for the order parameter and the critical surface, 
we prove that the critical behavior on complex networks
is in fact infinitely robust with respect to the presence of arbitrary short-loops. 
}

\begin{document}

\maketitle

\email{massimo.ostilli@roma1.infn.it}

\section{Introduction}
The simple but fundamental Curie-Weiss equation  
is the prototype of all mean-field models.
For an Ising model with a coupling $J/N$ defined over the fully connected graph of $N$ spins, 
the average magnetization $m$ satisfies the celebrated equation 
$m=\tanh(\beta Jm +\beta h)$, $\beta=1/T$ and $h$ being the inverse temperature and 
the external field. Similarly, Ising models defined on disordered graphs 
with a finite fixed (regular random graph) or random connectivity Poissonian distributed 
(classical random graph),
due to the tree-like nature
of the graphs, can be still exactly solved by using the Bethe-Peierls (BP) approach \cite{Mezard}, and the
resulting average magnetization $m$ obeys a more complicated mean field equation 
which, however, presents the same critical behavior of the Curie-Weiss equation with the classical
critical indices; in particular $m$ behaves as $m\sim \tau^{1/2}$, $\tau$ being the reduced temperature.
In the last 10 years, a strong renewed interest towards mean-field models
came from the study of cooperative phenomena defined on complex networks,
\textit{i.e.}, random graphs having a heterogeneous distribution of links among the sites \cite{DMAB}.
It turns out that, unlike the above random graphs, which are homogeneous, 
when the distribution of links is strongly heterogeneous, as happens when the degree
distribution $\mathcal{P}(k)$ of the graph follows a power law $\mathcal{P}(k)\sim k^{-\gamma}$,
with an exponent $\gamma<5$,
the average magnetization $m$ obeys an anomalous mean field equation giving rise 
to the critical behavior $m\sim \tau^{1/(\gamma-3)}$ for $\gamma>3$, or $m\sim T^{-1/(3-\gamma)}$ 
for $3>\gamma>2$ \cite{Review}.
As in the case of the homogeneous random graphs, also for the complex networks without short-range links,
the tool to achieve the mean-field equation for $m$ is basically the BP approach that is based on 
the tree-like assumption of the graph which, at least for $\gamma>3$, is locally true; 
in fact, if $c$ is the average connectivity, the shortest loops are of length $\mathop{O}(\log(N)/\log(c))$, 
so that in the thermodynamic limit there are no loops of finite length.
The analytical study of critical phenomena on complex networks has had a huge impact on
the understanding of collective behaviors such as diffusion, cooperation and percolation,  
in biological, social or technological contexts.
However, almost all these studies assume graphs locally tree-like 
(or at least hyper-graphs locally tree-like \cite{NewmanC})
which in realistic networks is almost never satisfied. 
Social networks, the neurons 
of the brain, the WWW and the Internet, 
are just a few examples in which the average clustering coefficient $C$ \cite{DMAB} is finite.
More precisely, whereas networks having a hierarchical structure usually share
a $k$ degree-dependent clustering coefficient of the form $C(k)\sim k^{-\alpha}$, with $\alpha \sim 1$, so that the
most connected (and most important) nodes are not clustered, there are other networks
having $C(k)\sim \mathop{O}(1)$ for almost any $k$. 
The former class includes \textit{e.g.} some social networks, language networks, 
the WWW, and the Internet at the autonomous system level, whereas the
latter class includes the Internet at the router level, the power grid, but also the brain.  
As discussed in \cite{Barabasi}, the reason for this difference is related to the fact that
in the second class wiring is expensive (economically or biologically) and the network, rather than 
hierarchically, is geographically organized. 
It is hence of fundamental importance to understand what is the role of a loopy topology in
complex networks from the point of view of collective behavior.
In particular: how does change the critical surface and - above all - 
are the analytical results accumulated over ten years of research in complex networks 
robust with respect to the presence of loops? And if yes, to what extent?
Notice that the crucial question concerns especially the loops of finite length but, 
due to the fact that the correlation length in these models remains finite,
we are not allowed to neglect these short-loops neither near the critical point, 
so that the above questions are quite far from being trivial.

At an intermediate level between loopy-like networks (\textit{e.g.} finite-dimensional lattices) 
and tree-like networks, 
stay the so called small-world networks \cite{Watts} in which we have an homogeneous distribution of links among the sites,
but also a finite clustering coefficient, so that tree-like based techniques as the BP cannot be used.
By using a completely different approach is however possible to solve exactly also these kind
of homogeneous small-world models at least in the paramagnetic phase (P) and to get their exact critical
surface and behavior \cite{SW}. If $J_0$ is the coupling associated to a given graph $(\mathcal{L}_0,\Gamma_0)$,
which in particular may have short-range links and short loops,
and $J$ is the coupling associated to a number $Nc/2$ of uniformly spread long-range links (so that they form
a classical random graph), 
then, for any $c>0$, the mean-field equation in these models is given by  
\begin{eqnarray}
\label{m0}
m=m_0(\beta J_0;c t m+\beta h), \quad t\defi \tanh(\beta J),
\end{eqnarray}
where $m_0(\beta J_0;\beta h)$ is defined as the average magnetization of the
model in the absence of the long-range connections, with a short-range coupling $J_0$ 
and in the presence of a generic external field $h$ at temperature $T=1/\beta$.
Eq. (\ref{m0}) is a natural  generalization of the Curie-Weiss mean-field equation. 
It is easy to check that however, for any $J_0\geq 0$, the critical behavior of Eq. (\ref{m0})
is classical, regardless of the local topology and clustering coefficient of the network (see also \cite{Hastings}).
In this Letter we face the natural extension of Eq. (\ref{m0}) towards a large class of 
heterogeneous small-world networks that can be considered as zero or, at most, weakly
degree-degree correlated.
After deriving the equation for the order parameter and the critical surface (thermal or bond-percolative),
to which we devote also a couple of explicit examples,
we prove that the critical behavior (thermal or bond-percolative) 
on these networks is never affected by the presence of 
a local non tree-like structure, provided the connectivity associated to such loopy structures
be in turns non heterogeneous. 

\label{models}
\section{Ising model on complex networks with loops}
The family of models we shall consider is built as follows.
Let $(\mathcal{L}_0,\Gamma_0)$ be any graph, 
$\mathcal{L}_0$ and $\Gamma_0$ being the set of vertices $i=1,\ldots,N$ and links, 
respectively.
Let us consider the Ising model defined on the graph $(\mathcal{L}_0,\Gamma_0)$ 
with a fixed coupling $J_0$ and in the presence of an arbitrary external field $\{h_i\}$ 
\begin{eqnarray}
\label{H0}
H_0= -J_{0}\sum_{(i,j)\in \Gamma_0}\sigma_{i}\sigma_{j}-\sum_i h_i\sigma_i.
\end{eqnarray}
We will call this \textit{the pure model}. 
Let us now consider the model obtained by removing randomly some links of the graph $(\mathcal{L}_0,\Gamma_0)$
and by adding new links as follows. Let us indicate with $c_{0;i,j}=0,1$ the adjacency matrix 
of the new graph in which some links of $\Gamma_0$ have been removed.
Given an ensemble $\mathcal{C}$ 
of random graphs $\bm{c}$, $\bm{c}\in\mathcal{C}$,
whose links are determined by the adjacency matrix elements $c_{i,j}=0,1$,
we define our \textit{heterogeneous small-world model} through the following Hamiltonian 
\begin{eqnarray}
\label{H}
H_{\bm{c}_0,\bm{c}}= -J_0\sum_{(i,j)\in \Gamma_0}c_{0;i,j}\sigma_{i}\sigma_{j}
 -h\sum_i \sigma_i
-J\sum_{i<j} c_{ij}\sigma_{i}\sigma_{j}.
\end{eqnarray}
The averages over the graph disorder (\ref{p0})-(\ref{hidden}) 
will be indicated as $\overline{\cdot}$. 
The variables
$c_{i,j}$ specify whether a ``long-range'' link (and then a coupling $J$) between the sites
$i$ and $j$ is present ($c_{i,j}=1$) or absent ($c_{i,j}=0$), whereas
the variables $c_{0;i,j}$ specify whether a ``short-range'' link $(i,j)\in\Gamma_0$
has been removed ($c_{0;i,j}=0$) or not ($c_{0;i,j}=1$).
Both $\{c_{0;i,j}\}$ and $\{c_{i,j}\}$ are assumed to be independent random variables with distributions
\begin{eqnarray}
\label{p0}
p_0(c_{0;i,j})=(1-p)\delta_{c_{0;i,j},1}+p\delta_{c_{0;i,j},0},
\end{eqnarray}
\begin{eqnarray}
\label{hidden}
 p_{ij}(c_{ij})=
f\left(p_i,p_j\right)\delta_{c_{ij},1}+
(1-f\left(p_i,p_j\right))\delta_{c_{ij},0},
\end{eqnarray}
where $p\in[0,1]$, and the $\{p_i\geq 0\}$ are a set of normalized hidden variables, $\sum_i p_i=1$ 
\cite{KimGoh0,CaldSatNewHidden},
each proportional to the average degrees $\{\bar{k}_i\}$ of the nodes $\mathcal{L}_0$ in the absence 
of the links $\Gamma_0$.
Usually, both $f(\cdot,\cdot)$ and the $\{p_i\}$ depend on one, or more, continuous parameters $\mu\in \mathcal{I}$, and on $N$.
Given the mean degree of $\bm{c}$, $c>0$: 
\begin{eqnarray}
\label{cave}
c=\sum_i \frac{\overline{k}_i}{N},
\end{eqnarray} 
we will assume that, for a continuous subset $\mathcal{J}\subset\mathcal{I}$, asymptotically in $N$ 
\begin{eqnarray}
\label{hidden1}
f\left(p_i,p_j\right)=cN p_ip_j. 
\end{eqnarray}
As a probability, Eq. (\ref{hidden1}) for $f(p_i,p_j)$ will be
manifestly violated in $\mathcal{I}\setminus\mathcal{J}$ whenever $cNp_ip_j>1$.
Note that, if $p_i\neq 0$ for any given $N$ (a requirement which is true for any graph
in which there are no isolated nodes), for $N\to\infty$ the terms $cN p_ip_j$
tends either to $0$ or to $\infty$, therefore, in $\mathcal{I}\setminus\mathcal{J}$,
the number of links $(i,j)$ for which
Eq. (\ref{hidden1}) is not true for $N$ large approaches 
\begin{eqnarray}
\label{hidden3}
\mathcal{N}_N\defi \sum_{i<j}\theta\left(cN p_ip_j-1\right),
\end{eqnarray}
where $\theta(x)=0$ or 1 if $x<0$ or $x\geq 0$, respectively.
If we parameterize $\mathcal{N}_N$ as $\mathcal{N}_N\sim N^\alpha$, 
and if the probability $\mathcal{P}(k)$ to have a vertex with degree $k$ scales, for $k$ large, 
as $\mathcal{P}(k)\sim k^{-\gamma}$, it is then easy to check that 
we have either $\alpha=0$ for $\gamma>3$, or $\alpha=3-\gamma$, for $\gamma\leq 3$.
As is intuitively clear (and as is possible to show rigorously \cite{SWSCL}), the free energy of (\ref{H}),
defined in the usual quenched way, has a leading $\mathop{O}(N)$ term coming from
the links satisfying Eq. (\ref{hidden1}), 
plus the $\mathop{O}(N^\alpha)$ rest. Therefore, in the thermodynamic limit, for $\gamma>2$, this rest 
becomes always negligible whereas, for $N$ large but finite, 
the results we below present are exact up to corrections per spin
which are near $0$ for $\gamma>3$, or $\mathop{O}(N^{2-\gamma})$ for $2<\gamma\leq 3$.

As an example one can consider the choice   
\begin{eqnarray}
\label{stat}
f\left(p_i,p_j\right)=1-e^{-cNp_ip_j}, \quad
 p_i=\frac{i^{-\mu}}{\sum_{j\in \mathcal{L}_0}j^{-\mu}},
\end{eqnarray}
where $\mu\in[0,1)$.
Eqs. (\ref{stat}) define the static model introduced in \cite{KimGoh0}.
Note that the ``fermionic'' constraint that avoids to have multiple bonds
in Eq. (\ref{hidden}) is implicit.
This constraint produces some weak dis-assortative degree-degree correlations for $\mu>1/2$ \cite{SatorrasHidden}.
In the thermodynamic limit $N\to\infty$, for $\mu\in (0,1)$, Eqs. (\ref{stat}) lead to a
number of long range connections per site distributed according
to a power law with mean $c$ and exponent $\gamma$ given by $\gamma=1+1/\mu$,
so that $\gamma\in(2,\infty)$. 
For $\mu\in(0,1/2)$ ($\gamma>3$) the first of Eqs. (\ref{stat}) takes the simpler form (\ref{hidden1}) 
while for $\mu\in[1/2,1)$ ($2<\gamma\leq 3$) it can be written as Eq. (\ref{hidden1})
only when $i$ and $j$ are sufficiently distant, $ij\gg N^{2-1/\mu}$, while for lower distances,
$ij\ll N^{2-1/\mu}$, we have $p_{ij}(c_{ij}=1)\simeq 1$. 

The class of our heterogeneous small-world models given by Eqs. (\ref{H})-(\ref{hidden}) with
the only conditions $c>0$ and $\gamma>2$ in $\mathcal{I}\setminus\mathcal{J}$ is very general.
Note in particular that the graph $(\mathcal{L}_0,\Gamma_0)$ is completely arbitrary 
and can contain closed paths of any length. We stress that the resulting
network, union of the graph $(\mathcal{L}_0,\Gamma_0)$ in which each link is removed
with probability $p$, with the random graph $\bm{c}$, can be seen as a gran-canonical heterogeneous generalization
of the original small-world graph of Watts and Strogatz \cite{Watts}, though we here do not have a true
rewiring. Since we let the probability $p\in[0,1]$ and the mean $c\in(0,\infty)$ arbitrary, our way
to build small-world networks is more general even for the homogeneous case $p_i\equiv 1/N$.
However, we can always restrict our class of small-world networks to the ones having a total
average connectivity which does not change with $p$.
To this aim we can choose $c$ such that the number of links of the graph $\bm{c}$
is equal to the number of removed links of $(\mathcal{L}_0,\Gamma_0)$.
Up to corrections $\mathop{O}(1/\sqrt{N})$ we can accomplish this for any sample
by simply taking $c=c_0 p$, where $c_0$ is the average connectivity of $(\mathcal{L}_0,\Gamma_0)$.

\section{Self consistent equation}
In the following, we will use the label $\mathop{}_0$ 
to specify that we are referring
to the pure model with Hamiltonian (\ref{H0}).
Let $m_{0i}(\beta J_0,\{\beta h_j\})$ be the stable magnetization 
of the spin $i$ of the pure model (\ref{H0}) 
at inverse temperature $\beta$. 
Then the order parameter $m$ of the model (\ref{H})-(\ref{hidden1}) obeys
\begin{eqnarray}
\label{THEOa}
&& m=\sum_i m_{0i}(\beta J_0^{(\mathrm{F})};
\{Np_jctm+\beta h\})p_i,\\
&&t\defi\tanh(\beta J), \nonumber \\
&&t_0\defi \tanh(\beta J_0^{(\mathrm{F})})\defi\tanh(\beta J_0)(1-p). \nonumber
\end{eqnarray}
The meaning of the order parameter $m$ is quite natural:
\begin{eqnarray}
\label{orderpar}
m=\sum_i p_i \overline{\media{\sigma_{i}}},
\end{eqnarray} 
($\media{\cdot}$ thermal average)
while for any correlation function ${{C}}$, and for some positive $\delta$, we have
\begin{eqnarray}
\label{THEOh}
{{C}}&=&
{{C}}_0(\beta J_0^{(\mathrm{F})};\{Np_jctm+\beta h\})+
\mathop{O}\left({N^{-\delta}}\right),
\end{eqnarray} 
where ${{C}}_0(\beta J_0,\{\beta h_i\})$ is the correlation function of the 
pure model (\ref{H0}) at inverse temperature $\beta$. 
The proof of (\ref{THEOa})-(\ref{THEOh}) is almost identical to the
proof of Eq. (\ref{m0}) \cite{SW}, the only difference being related to the fact that,
here, the probability $p_{ij}(c_{ij})$ depends explicitly on the link $(i,j)$.
As Eq. (\ref{m0}), also Eqs. (\ref{THEOa})-(\ref{THEOh}) are exact in the P region and provide the
exact critical surface and critical exponents, while $m$ is correct up to $\mathop{O}(\overline{k}/\overline{k^2})$ 
corrections out of the P region. A detailed proof will be reported~elsewhere \cite{SWSCL}. 

As an immediate consequence of Eq. (\ref{THEOa}) we get the susceptibility 
$\tilde{\chi}$ ($=\partial m/\partial(\beta h)$) of the random model
\begin{eqnarray}
\label{THEOchie}
\tilde{\chi}
= \frac{\sum_ip_i\sum_j\tilde{\chi}_{0;i,j}\left(\beta J_0^{(\mathrm{F})};\{Np_lctm+\beta h\}\right)}
{1-ctN\sum_{i,j}\tilde{\chi}_{0;i,j}
\left(\beta J_0^{(\mathrm{F})};\{Np_lctm+\beta h\}\right)p_ip_j},
\end{eqnarray}
where $\tilde{\chi}_{0;i,j}(\beta J_0;\{\beta h_j\})$ stands for the two-points connected correlation function
of the pure model (\ref{H0}) at inverse temperature $\beta$:
$\tilde{\chi}_{0;i,j}\defi 
\media{\sigma_i\sigma_j}_0-\media{\sigma_i}_0\media{\sigma_j}_0$.
Note that, even if the model (\ref{H0}) with $h_i\equiv h$
is translational invariant, 
the model (\ref{H})-(\ref{hidden}) is no more translational invariant (except the case $p_i\equiv 1/N$).
Note in particular that $\tilde{\chi}$ is not proportional
to the sum of the connected correlation functions.
In fact, from Eqs. (\ref{orderpar}) and (\ref{THEOh}) it follows
\begin{eqnarray}
\label{THEOchie1}
\tilde{\chi}= \sum_{i,j}p_i\left[\overline{\media{\sigma_i\sigma_j}-
\media{\sigma_i}\media{\sigma_j}}\right].
\end{eqnarray}

\section{Critical surface (thermal and percolative)}
From the self-consistent Eq. (\ref{THEOa}) we see that, in the thermodynamic limit, 
the critical surface $\beta_c$ satisfies
\begin{eqnarray}
\label{THEOcrit}
ct_cN \sum_{i,j}\tilde{\chi}_{0;i,j}
\left(\beta_c J_0^{(\mathrm{F})};0\right)p_ip_j=1. 
\end{eqnarray}
For any given set ($J_0$, $J$, $p$, $\{p_i\}$),
Eq. (\ref{THEOcrit}) gives the critical surface of the model in the plane $(\beta,c)$. 
Furthermore, from Eq. (\ref{THEOa}) it is immediate to recognize that 
$\beta_c \leq \beta_{c0}^{(\mathrm{F})}$, with $\beta_c =\beta_{c0}^{(\mathrm{F})}$ only for $J_0=0$,
where $\beta_{c0}^{(\mathrm{F})}$ is the critical temperature of the pure model (\ref{H0})
but with coupling $J_0^{(\mathrm{F})}$. 

The theory can be projected toward the limit $\beta\to\infty$ where
we get an effective percolation theory.
Here the region P corresponds to the region in which,
in the thermodynamic limit, the parameters $(c,c_0,p)$
are such that no giant connected component exists ($m=0$).
Note in particular that, if $c_{0c}$ is the percolation threshold of 
the pure graph $(\mathcal{L}_0,\Gamma_0)$ (if $c_{0c}$ does not exist we can set formally $c_{0c}=\infty$)
in order to remain in such a region, 
the connectivity $c_0^{(p)}=c_0(1-p)$ of the graph obtained from the graph
$(\mathcal{L}_0,\Gamma_0)$ in which each link has been removed at random with probability $p$,
must satisfy $c_0(1-p)\leq c_{0c}$, otherwise a giant connected component already exists.
From Eq. (\ref{THEOcrit}), for $c_0(1-p)\leq c_{0c}$, 
the equation for the percolation threshold $c_c$ as a function of $p$ is given by
\begin{eqnarray}
\label{THEOcrit5a}
&& c_cN \sum_{i,j}\tilde{\chi}_{0;i,j}\left(\tanh^{-1}(1-p);0\right)p_ip_j=1,
\end{eqnarray}
where we have used the fact that $\lim_{\beta\to\infty}\tanh(\beta J_0^{(\mathrm{F})})=\tanh(1-p)$.
Alternatively, Eq. (\ref{THEOcrit5a}) can be rewritten in terms of only graph elements as
\begin{eqnarray}
\label{THEOcrit5}
c_cN \sum_{i,j}\left(\delta_{i,j}+\mathcal{N}_{0;i,j}^{(p)}\right)p_ip_j=1, \quad c_0(1-p)\leq c_{0c},
\end{eqnarray}
where $\mathcal{N}_{0;i,j}^{(p)}=1$ if, in the graph $(\mathcal{L}_0,\Gamma_0)$
from which each link has been removed at random with probability $p$,
between the vertex $i$ and the vertex $j$ there exists at least a path of links, 
and $\mathcal{N}_{0;i,j}^{(p)}=0$ otherwise.

Given $p$, if the condition $\quad c_0(1-p)\leq c_{0c}$ is not satisfied, then
a giant connected component is present and we can set $c_c=0$.
It is interesting to see in more details the case in which we choose $c=c_0p$ so
that, as we vary $p$, the total connectivity is fixed and equal to $c_0$ (as in a rewired small-world).
This study is important since it leads us to understand how the presence of short loops affects
diffusion processes on general networks. In particular, a strong interest
regards the question: \textit{``in the presence of short loops how does 
the percolation threshold change?''} 
If we set $c=c_0p$, from Eq. (\ref{THEOcrit5}) we get the percolation threshold $c_{0c}$ as
a function of the rewiring parameter $p$
\begin{eqnarray}
\label{THEOcrit5b}
c_{0c}^{(p)}pN \sum_{i,j}\left(\delta_{i,j}+\mathcal{N}_{0;i,j}^{(p)}\right)p_ip_j=1, \quad c_0^{(p)}(1-p)\leq c_{0c}.
\end{eqnarray}
From Eq. (\ref{THEOcrit5b}) we see that $p$ may have two effects on $c_{0c}^{(p)}$: one tends to
decrease $c_{0c}^{(p)}$ while the other tends to increase $c_{0c}^{(p)}$. 
In general $c_{0c}^{(p)}$ decreases with $p$ for $p$ small, while grows
for $p$ larger. Let us consider the case in which $p_i\equiv 1$;
\textit{i.e.}, the classical small-world (no heterogeneity).
In this case Eq. (\ref{THEOcrit5b}) simplifies as
\begin{eqnarray}
\label{THEOcrit5c}
c_{0c}^{(p)}p\tilde{\chi}_0(\tanh^{-1}(1-p);0)=1, \quad c_0^{(p)}(1-p)\leq c_{0c}.
\end{eqnarray}
For example, if $(\mathcal{L}_0,\Gamma_0)$ is the 
Erd$\mathrm{\ddot{o}}$s-R$\mathrm{\acute{e}}$nyi graph (in the canonical representation),
with mean connectivity $c_{0}$, by using in Eq. (\ref{THEOcrit5c})  
$\tilde{\chi}_0(\beta J_0;0)=\left[1-c_0\tanh(\beta J_0)\right]^{-1}$ (valid in the P$_0$ region)
we get back obviously the well known percolation threshold $c_{0c}^{(p)}=1$, independently of the value of $p$.  
Depending on the problem, given $c_0<c_{0c}$, in general one can be more interested in reading Eq. (\ref{THEOcrit5})
either as an equation for $p$ or for $c$. 
We will see later a simple example,
while further analysis and applications to the percolation problem will be reported elsewhere \cite{SWSCL}.

\section{Critical behavior}
We prove now that, if $c>0$ 
and $(\mathcal{L}_0,\Gamma_0)$ is not a heterogeneous graph,
then the critical behavior of the model (\ref{H})-(\ref{hidden1})
does not depend either on $(\mathcal{L}_0,\Gamma_0)$ or on $(J_0,J,p,c)$.
In \cite{SW}, we have shown this result for the homogeneous small-world case
$p_i\equiv1/N$ and with $p=0$. More precisely in such a case the critical behavior 
has been shown to be classical mean field for $t_0^{}\geq 0$, while
for $t_0^{}< 0$ first-order phase transitions are also possible (see also \cite{SWMC}).
Here we will restrict the analysis only to the case $t_0^{}\geq 0$.  
First of all from Eqs. (\ref{THEOchie}) and (\ref{THEOcrit}) we observe immediately that  
the critical exponents for the susceptibility, above and below the critical temperature, 
are both equal to 1. 
Let us now turn to the analysis of the order parameter near the critical point.
In the following we set $h=0$ and 
it is understood that we consider only the positive solution of (\ref{THEOa}) $m\geq 0$.
For $J_0=0$ we~have
\begin{eqnarray}
\label{THEOp}
m_{0i}(\beta J_0^{(\mathrm{F})};\{\beta h_j\})=
\tanh(\beta h_i)
\end{eqnarray} 
so that Eq. (\ref{THEOa}) strongly simplifies in
\begin{eqnarray}
\label{THEOp1}
m^{}=g(m^{}), \quad 
g(m^{})\defi \sum_i \tanh(Np_ict^{}m^{})p_i .
\end{eqnarray}
The critical behavior of the model for $J_0=0$, 
has been already studied  in \cite{KimStat}. Eq. (\ref{THEOp1}) is equal to Eq. (21) of \cite{KimStat}.
We recall that, if the distribution $\{p_j\}$ has a power-law character, we cannot derive the correct
critical behavior of the system by simply expanding the sum $g(m^{})$ term by term for small $m^{}$. 
As shown in \cite{KimStat}, it is necessary to keep track of all the terms of the sum $g(m^{})$. 
This is done by evaluating the sum with the corresponding integral
which gives rise to a singular term proportional to $(m^{})^{\gamma-2}$ plus regular terms
proportional to $(m^{})$, $(m^{})^{3}$ and so on, where $\gamma=1+1/\mu$, if $\mu$ is the power law
exponent associated to $\{p_j\}$.
As a consequence, when we solve Eq. (\ref{THEOp1}) to leading order in $m^{}$, 
if $T$ and $\tau$ indicate the temperature and the reduced temperature, respectively,
we get the well known anomalous mean-field behavior: 
$m^{}\sim\mathop{O}(\tau^{1/2})$ (\textit{i.e.} classical mean-field) for $\gamma>5$,  
$m^{}\sim\mathop{O}(\tau^{1/(\gamma-3)})$ for $3<\gamma<5$, and  
$m^{}\sim\mathop{O}(T^{-(\gamma-2)/(3-\gamma)})$ for $2<\gamma<3$.  
Note that the critical behavior of the order parameter $m^{}=\sum_i\overline{\mediaT{\sigma_i}}p_i$ 
is different from the unweighted 
one defined as $\overline{m}=\sum_i\overline{\mediaT{\sigma_i}}/N$ when $2<\gamma<3$. In such a case,
from Eq. (\ref{THEOh}) one can use 
$\overline{m}\sim t^{}m^{}$ from which it follows that 
$\overline{m}\sim\mathop{O}(T^{-1/(3-\gamma)})$ for $2<\gamma<3$. 

Let now be $J_0>0$. 
It is clear that if the graph $(\mathcal{L}_0,\Gamma_0)$ is in turn
a pure scale free graph with exponent $\gamma'$, then the joined network
will have an anomalous critical behavior characterized by the minimum between
$\gamma$ and $\gamma'$. Less obvious is to understand what happens if $(\mathcal{L}_0,\Gamma_0)$
has a finite dimensional structure or some special topology with short loops.
In particular we can pose the question: \textit{``does the critical behavior change
by adding, via short loops, many paths between far spins, or may the critical exponent for $m$ depend on $J_0$?}'' 
Let us consider the self-consistent equation (\ref{THEOa}) in general.     
The exact expression of $m_{0i}(\beta J_0^{(\mathrm{F})};\{\beta h_j\})$ for a generic
non homogeneous external field $\{h_j\}$ represents a formidable task. 
Note that, as above mentioned, to analyze the critical behavior we cannot expand for small fields $\{\beta h_j\}$.
However, since we have proved that, for $J_0^{(\mathrm{F})}>0$, $\beta_{c0}^{(\mathrm{F})}>\beta_c$,
in the region $\beta_{c0}^{(\mathrm{F})}>\beta>\beta_c$ we can
perform an high temperature expansion of $m_{0i}(\beta J_0^{(\mathrm{F})};\{\beta h_j\})$ 
in the parameter $t_0^{}=\tanh(\beta J_0^{(\mathrm{F})})$.
It is then easy to see that
\begin{eqnarray}
\label{THEOp2}
&& m_{0i}(\beta J_0^{(\mathrm{F})};\{\beta h_j\})=\tanh(\beta h_i) ~~~~~~~~~~~\nonumber \\
&& + t_0^{}\left[1-\tanh^2(\beta h_i)\right]
\sum_{j\in\mathcal{N}_0(i)}\tanh(\beta h_j)+\mathop{O}(t_0^2),~~~~~~
\end{eqnarray} 
where $\mathcal{N}_0(i)$ is the set of the first neighbors of the vertex $i$ in the graph $(\mathcal{L}_0,\Gamma_0)$.
Plugging Eq. (\ref{THEOp2}) into Eq. (\ref{THEOa})
\begin{eqnarray}
\label{THEOp3}
&& m^{}=g(m^{})+\Delta_1(m^{})+\Delta_2(m^{}),
\end{eqnarray}
where we have introduced
\begin{eqnarray}
\label{THEOp4}
\Delta_1(m^{})\defi 
t_0^{}\sum_i p_i\sum_{j\in\mathcal{N}_0(i)}\tanh(Np_jct^{}m^{}),
\end{eqnarray}
and $\Delta_2(m^{})$ is the rest coming from Eq. (\ref{THEOp2})
which, near the critical point, $|\Delta_2(m^{})|\ll\Delta_1(m^{})$.
Let us compare the bigger contribution $\Delta_1(m^{})$ with the ``pure $J_0=0$ term'' $g(m^{})$  and
let us focus on the simpler cases in which the graph $(\mathcal{L}_0,\Gamma_0)$
has a fixed connectivity $|\mathcal{N}_0(i)|\equiv c_0$. 
Let us suppose first that $c_0=1$ (\textit{i.e.}, $(\mathcal{L}_0,\Gamma_0)$ is an ensemble of dimers).
In general, given any normalized distribution $p_i\geq 0$, and any function $f(x)\geq 0$
increasing with $x$, the following strict (for $p_i\neq 1/N$) inequality holds 
\begin{eqnarray}
\label{THEOp5}
\sum_i p_i f(p_{j_0(i)})<\sum_i p_i f(p_i), 
\end{eqnarray}
where $j_0(i)$ stands for the first single neighbor of $i$ in $(\mathcal{L}_0,\Gamma_0)$.
Note that $i\rightarrow j_0(i)$ is a bijection on $\mathcal{L}_0$ and that
$j_0(i)\neq i$. We can however formally enlarge the definition of $j_0(i)$ 
to include also the case $j_0(i)=i$.
The inequality (\ref{THEOp5}) tells us that when we choose $j_0(i)=i$ we get an optimal
overlap between the distribution $\{p_i\}$ and the function $f(\cdot)$. 
For the general case $|\mathcal{N}_0(i)|\equiv c_0\geq 1$, for any $i$, we can
enumerate the $c_0$ neighbors of $i$ in the same way 
as $j_0^{(1)}(i),\ldots,j_0^{(c_0)}(i)$, so that for each upper index $l$
(that represents an oriented direction) $j_0^{(l)}(i)$ is a bijection on $\mathcal{L}_0$ on which 
we can apply (\ref{THEOp5}) to get finally 
\begin{eqnarray}
\label{THEOp6}\nonumber
\sum_i p_i \sum_{j\in\mathcal{N}_0(i)} f(p_j)=
\sum_{l=1}^{c_0}\sum_i p_i f(p_{j_0^{(l)}(i)})<c_0\sum_i p_i f(p_i).  
\end{eqnarray}
By applying this equation to our case with $f(x)=\tanh(x)$ and for $t_0^{}>0$
we see that for any $m^{}>0$ 
\begin{eqnarray}
\label{THEOp7}
0< |\Delta_2(m^{})|<\Delta_1(m^{})<t_0^{}c_0 g(m^{}).
\end{eqnarray}
In conclusion from Eqs. (\ref{THEOa}) and (\ref{THEOp7}) we get
\begin{eqnarray}
\label{THEOp8}\nonumber
g(m^{})< m^{}<\left(1+t_0^{}c_0 \right)g(m^{})+\mathop{O}(t_0^2)g(m^{})+\mathop{o}(g(m^{})),
\end{eqnarray}
where $\mathop{o}(x)$ is such that $\lim_{x\to 0}\mathop{o}(x)/x=0$.
In general, for $t_0$ finite, it is possible to prove that \cite{SWSCL}
\begin{eqnarray}
\label{THEOp8g}
g(m^{})< m^{}<\tilde{\chi}_0\left(\beta J_0^{(\mathrm{F})};0\right)g(m^{})+\mathop{o}(g(m^{})),
\end{eqnarray} 
where $\tilde{\chi}_0(\beta J_0^{(\mathrm{F})};0)$ is the susceptibility of
the pure model (\ref{H0}) with coupling $J_0^{(\mathrm{F})}$ and $h_i\equiv 0$. 
Since near the critical point, in the region $\beta_{c0}^{(F)}>\beta>\beta_c$, it is
$\tilde{\chi}_0(\beta J_0^{(\mathrm{F})};0)<\infty$, we see that
Eq. (\ref{THEOp8g}) implies that the critical behavior of Eq. (\ref{THEOa})
remains always that 
corresponding to the term $g(m^{})$, \textit{i.e.} as if it were $J_0=0$.
Hence, \textit{g.e.}, if $(\mathcal{L}_0,\Gamma_0)$ is a $d_0$-dimensional lattice and 
$\{p_i\}$ generates a scale-free graph with degree exponent $3<\gamma<5$, Eq. (\ref{THEOp8g}) 
tells us that near the critical temperature for the model (\ref{H}) $m\sim \tau^{1/(\gamma-3)}$, 
regardless the values of $J_0,J,p,c$ and $d_0$. 
We can also consider the case in which $(\mathcal{L}_0,\Gamma_0)$ is a
Poissonian graph 
with mean connectivity $c_0$. To this aim
we can start from the fully connected graph and remove from it randomly
each of its $N(N-1)/2$ links with probability $p=1-c_0/N$; the resulting graph
will be our Poissonian graph with mean connectivity $c_0$. 
Since we have already proved
that when $(\mathcal{L}_0,\Gamma_0)$ is the fully connected graph 
with a couplings $\mathop{O}(1/N)$ the critical behavior 
remains equal to that of the model with $J_0=0$ (the effective
coupling in this case being $\beta J_0^{(\mathrm{F})}=\tanh(\beta J_0)c_0/N$), 
we conclude that, also when $(\mathcal{L}_0,\Gamma_0)$ is a Poissonian graph, the critical behavior
on the total graph remains the same as if it were~$J_0=0$.

\section{Examples}
Concerning the critical surface, particularly simple are the cases in which $(\mathcal{L}_0,\Gamma_0)$ is a set
of disconnected finite clusters (points, dimers, triangles, \dots, or mixtures of them).
Note that for such $(\mathcal{L}_0,\Gamma_0)$ there is no percolation threshold (or formally $c_{0c}=\infty$).

Let $(\mathcal{L}_0,\Gamma_0)$ be a set of $N/2$ disconnected dimers.
This case represents the simplest example with $J_0\neq 0$ in which 
$m_{0i}(\beta J_0^{(\mathrm{F})};\{\beta h_j\})$ can be exactly calculated.
We have
\begin{eqnarray}
\label{THEOdim}
m_{0i}(\beta J_0^{(\mathrm{F})};\{\beta h_j\})=\frac{\tanh(\beta h_i)+t_0\tanh(\beta h_{j_0(i)})}
{1+t_0\tanh(\beta h_i)\tanh(\beta h_{j_0(i)})},
\end{eqnarray} 
where $t_0=\tanh(\beta J_0^{(\mathrm{F})})$ and $j_0(i)$ stands for the first neighbor of $i$.
By derivation we get the correlation function 
$\tilde{\chi}_{0;i,j}$ which, in the P region, takes the form
\begin{eqnarray}
\label{THEOdim1}
\tilde{\chi}_{0;i,j}(\beta J_0^{(\mathrm{F})};0)=\delta_{i,j}+t_0~\delta_{i,j_0(i)}.
\end{eqnarray} 
Therefore for the critical surface we have
\begin{eqnarray}
\label{THEOdim2}
ct_c N \left[\sum_i p_i^2+t_0\sum_i p_ip_{j_0(i)}\right]=1.
\end{eqnarray}
With respect to the critical surface of the model with $J_0=0$ 
we see in Eq. (\ref{THEOdim2}) the presence of a new term proportional to $t_0$.
How much this term affects $t_c$ depends on how the dimers are placed,
\textit{i.e.}, on how we choose the first neighbors $\{j_0(i)\}$.
Since by definition the dimers are not connected, in general for $j_0(i)$ we can
take $j_0(i)=i+k, \mod N$ where $k$ is a constant integer in the range $[1,N]$.
The exact evaluation of $t_c$ for $N$ large remains simple only if
$k$ does not depend on $N$ or $k=\mathop{O}(N)$. 
Under the choice (\ref{stat}) for the former case we have 
\begin{eqnarray}
\label{THEOdim3}
ct_c{(1+t_{c0})(1-\mu)^2}(1-N^{2\mu-1})(1-2\mu)^{-1}=1,
\end{eqnarray}     
whereas for the latter the term proportional to $t_0$ 
in the thermodynamic limit becomes negligible. 
Of course, we re-find that for $\mu>1/2$ (\textit{i.e.} $\gamma<3$) $t_c\to 0^+$.

Let $(\mathcal{L}_0,\Gamma_0)$ be now a set of $N/3$ disconnected triangles.
For simplicity here we assume $\mu=0$, \textit{i.e.}, the graph disorder is just
the classical one (Erd$\mathrm{\ddot{o}}$s-R$\mathrm{\acute{e}}$nyi graph) and we calculate
the percolation threshold for the problem with fixed total connectivity that
we have under the choice $c=c_0p$, with $c_0=2$.
For the susceptibility of the pure model with generic coupling $\beta J_0$ we get 
\begin{eqnarray}
\label{THEOcrit5fc}
\tilde{\chi}_0(\beta J_0;0)=\frac{3e^{3\beta J_0}+e^{-\beta J_0}}{e^{3\beta J_0}+3e^{-\beta J_0}}.
\end{eqnarray}
From Eq. (\ref{THEOcrit5c}) and (\ref{THEOcrit5fc}), by using the replacement $\beta J_0\to\tanh^{-1}(1-p)$,
we get the equation for the percolation threshold $p_c$ (which is solved for $p_c=0.183406$)   
\begin{eqnarray}
\label{THEOcrit5fc1}
2p\frac{3e^{3\tanh^{-1}(1-p)}+e^{-\tanh^{-1}(1-p)}}{e^{3\tanh^{-1}(1-p)}+3e^{-\tanh^{-1}(1-p)}}=1.
\end{eqnarray}

In the above examples we had $d_0=0$. For a chain of $N$ spins
we can evaluate $T_c$ from Eq. (\ref{THEOcrit}) by using $\tilde{\chi}_{0;i,j}=t_0^{|i-j|}$.
In Fig. (1) we plot simulations for the susceptibility $\chi$ as a function of $T$ for several 
system sizes and compare the location of the maximums with the theoretical $T_c$.
\begin{figure}[tbh]
\epsfxsize=67mm \centerline{\epsffile{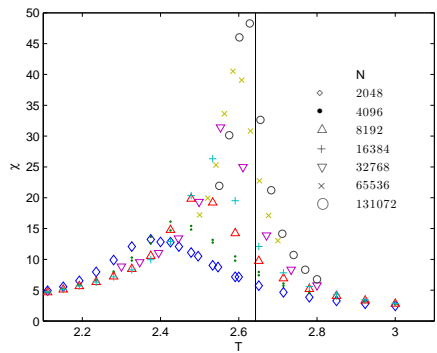}}
\caption{Plots of $\chi$ as a function of $T$ for a chain of N spins in networks
generated via Eqs. (\ref{stat}) with $c=1$, $J=J_0=1$, $\mu=1/3$ (corresponding to $\gamma=4$), and $p=0$. 
The vertical line comes from the solution of Eq. (\ref{THEOcrit}) with $N=131072$.}
\end{figure}

\section{Conclusions}
The results presented in this Letter are easily generalized to the case in which
$J_0$ and $J$ are random couplings and in particular we find that also for the
spin-glass critical behavior a similar robustness theorem applies.
These, as well as other considerations concerning 
the existence of strong finite size effects (see Eq. (\ref{THEOh}))
will be reported elsewhere \cite{SWSCL}.    
Our analysis sheds also some light to the issues related to complex networks 
embedded in a metric space. Although we here do not give the long-range couplings
$J$ a metric structure, the graph $(\mathcal{L}_0,\Gamma_0)$ 
is completely general so that it can be also a metric graph
equipped with an arbitrary distance $||i-j||_{_{_0}}$, which is the natural distance
through which the correlation functions via Eq. (\ref{THEOh}) are expressed.
An interesting generalization of the Ginzburg criterion
for the relevance of critical fluctuations has been recently derived 
for Ising models built on annealed networks embedded in a metric space \cite{Ginestra}.    
We recall however that the Ginsburg criterion alone, when fulfilled, does not provide
the critical exponent of the order parameter $m$,
especially if a heterogeneous distribution of links, and then a continuous
dependence on the parameters of this distribution, is considered.
At the heart of our robustness theorem (\ref{THEOp8}) lies the general inequality
(\ref{THEOp5}) whose lhs and rhs 
corresponds
to the short-range and to the long-range couplings of the model,  respectively.
We see then that the existence of long-range couplings in a model
makes its order parameter already optimized with respect to the addition of any possible regular,
or weakly random (\textit{i.e.} Poissonian like), structure; 
only the addition of further long-range links sufficiently 
heterogeneous may change the critical behavior of the system.
This work was supported by SOCIALNETS. We thank A. L. Ferreira for 
many useful discussion.


\end{document}